\documentclass{emulateapj}
\usepackage{natbib}
\usepackage{apjfonts}
\usepackage{lscape}
\usepackage{multirow}

\newcommand{\rl}{$R_{\rm BLR} - L$}
\newcommand{\msigma}{$M_{\rm BH}-\sigma_{\star}$}
\newcommand{\ml}{$M_{\rm BH}-L_{\rm bulge}$}
\newcommand{\mbh}{$M_{\rm BH}$}
\newcommand{\sersic}{S\'{e}rsic}

\shorttitle{Black Hole Mass of NGC\,5273}
\shortauthors{Bentz et al.}

\received{}
\accepted{}

\begin{document}

\title{The Mass of the Central Black Hole in the Nearby Seyfert Galaxy NGC\,5273}

\author{ Misty~C.~Bentz\altaffilmark{1},
Daniel~Horenstein\altaffilmark{1},
Craig~Bazhaw\altaffilmark{1},
Emily~R.~Manne-Nicholas\altaffilmark{1},
Benjamin~J.~Ou-Yang\altaffilmark{1},
Matthew~Anderson\altaffilmark{1},
Jeremy~Jones\altaffilmark{1},
Ryan~P.~Norris\altaffilmark{1},
J.~Robert~Parks\altaffilmark{1},
Dicy~Saylor\altaffilmark{1},
Katherine~G.~Teems\altaffilmark{1},
Clay~Turner\altaffilmark{1}
}

\altaffiltext{1}{Department of Physics and Astronomy,
		 Georgia State University,
                 25 Park Place, Suite 600,
		 Atlanta, GA 30303, USA;
		 bentz@astro.gsu.edu}

\begin{abstract}
We present the results of a reverberation-mapping program targeting
NGC\,5273, a nearby early-type galaxy with a broad-lined active
galactic nucleus.  Over the course of the monitoring program,
NGC\,5273 showed strong variability that allowed us to measure time
delays in the responses of the broad optical recombination lines to
changes in the continuum flux.  A weighted average of these
measurements results in a black hole mass determination of $M_{\rm BH}
= (4.7 \pm 1.6) \times 10^6$\,M$_{\odot}$.  An estimate of the size of
the black hole sphere of influence in NGC\,5273 puts it just at the
limit of the resolution achievable with current ground-based large
aperture telescopes.  NGC\,5273 is therefore an important future
target for a black hole mass determination from stellar dynamical
modeling, especially because it is the only nearby early-type galaxy
hosting an AGN with a reverberation-based mass, allowing the best
comparison for the masses determined from these two techniques.
\end{abstract}

\keywords{galaxies: active --- galaxies: nuclei --- galaxies: Seyfert}

\section{Introduction}

Over the last $\sim$25 years, observational and computational studies
have led to the widely accepted belief that supermassive black holes
(black holes with masses $M_{\rm BH} = 10^6 - 10^{10}$\,M$_{\odot}$)
play a significant role in galaxy evolution and cosmology.  To truly
understand the physical processes at play in an active galactic
nucleus (AGN) and the effects on the host galaxy and its surroundings,
it is necessary to know the mass of the central black hole that is
involved.

Currently, it is only feasible to directly measure black hole masses
in relatively nearby galaxies through dynamical modeling (which is
limited by spatial resolution) or reverberation mapping (which is
limited by time).  
%
%
For quiescent galaxies, \mbh\ is derived from modeling the dynamics of
stars and/or gas on parsec scales in a galaxy nucleus (e.g.,
\citealt{gultekin09,mcconnell13}).  For AGNs, the response of
milliparsec-scale gas in the broad line region (BLR) to the accretion
disk flux variability probes the gravitational force of the black hole
in a technique known as reverberation mapping (e.g.,
\citealt{peterson04}).  Dynamical studies rely on spatial resolution
and are therefore limited to the most nearby galaxies ($\lesssim
100-150$\,Mpc).  Reverberation mapping relies on temporal resolution
and is not inherently distance limited, but timing arguments and
resource availability have generally limited reverberation studies to
$z<0.1$ ($\sim 400$\,Mpc).  The \mbh\ measurements that result from
directly probing the gravity of the supermassive black hole through
these techniques provide the foundation for all black hole scaling
relations used to investigate \mbh\ at cosmological distances. As
such, the \mbh\ estimates provided by the scaling relations are
directly affected by the quality of the observations and measurements
used to calibrate the relationships.

Both the dynamical and reverberation masses currently suffer from
several inherent uncertainties and potential systematic biases (for
recent discussions, see \citealt{graham11} for the dynamical studies
and \citealt{peterson10} for the reverberation studies).  One of the
most pernicious uncertainties affecting reverberation masses is the
exact translation from reverberation measurements to calibrated black
hole mass.  The so-called ``f factor'' is an order unity scaling
factor that depends on the detailed geometrical and dynamical
properties of the broad-line region gas that is used to probe the
gravitational influence of the black hole.  Recent progress has shown
promise in constraining the ``f factor'' for individual reverberation
targets with high-quality datasets \citep{pancoast14}.  

What is currently lacking, however, is a set of black holes with both
dynamical and reverberation masses that can be directly compared to
ensure that all black hole masses are on the same mass scale.  Only
three AGNs in the current reverberation sample are close enough and
have massive enough black holes that they can be investigated through
stellar dynamical modeling --- NGC\,3227 \citep{davies06}, NGC\,4151
\citep{onken04,onken14}, and NGC\,6814 (\citealt{bentz09c} for the
reverberation mass; observations of the dynamics in the galaxy nucleus
are in progress at this time).  Comparison of black hole masses
derived from independent techniques in the same galaxies is an
important cross check, because these black hole masses are the
foundation upon which all scaling relationships are built.  Stellar
dynamical modeling, in particular, is often described as the ``gold
standard'' of supermassive black hole masses.  However, all stellar
dynamical masses result from axisymmetric modeling codes that are
unable to account for the unique dynamics that arise in the presence
of a bar.  NGC\,3227, NGC\,4151, and NGC\,6814 are all late-type
spiral galaxies with bars.  Recently, \citet{onken14} showed that even
the weak bar in NGC\,4151 can induce a strong bias in the resultant
black hole mass from stellar dynamical modeling, causing the black
hole mass to be overestimated.  This bias would not be expected to
affect the reverberation masses, where the tracer particles or ``test
masses'' are located much deeper in the gravitational potential well
of the black hole, at distances of only a few light days ($\sim
0.01$\,pc for nearby Seyferts), where they would be unaffected by the
dynamics of a galaxy-scale bar.

Motivated by the small number of AGNs with reverberation masses that
are also suitable targets for dynamical studies, we have begun a
program to identify and monitor all AGNs that may be suitable for both
reverberation and dynamical studies.  The rarity of broad-lined AGNs
in the local Universe is the fundamental limitation in this endeavor,
as this rarity generally translates to a large distance and therefore
an inability to spatially resolve the galaxy nucleus on the scales
necessary for dynamical modeling.  Our goal is to build the largest
sample possible, although the sample is not likely to be larger than
$\sim 10$ AGNs given the current technological constraints.  In this
manuscript, we present a reverberation-based mass for the central
black hole in NGC\,5273, the first reverberation target in an
early-type unbarred galaxy at a distance suitable for ground-based
dynamical studies.  In $\S 2$, we describe the observations, and in
$\S 3$, the light curve analysis. The black hole mass determination is
described in $\S 4$.  Finally, we discuss the results of our analysis
and summarize our findings in $\S 5$ and $\S 6$.

\section{Observations}

NGC\,5273 is a broad-lined Seyfert located at ${\rm RA} = 13{\rm h}
42{\rm m} 08.3{\rm s}$, ${\rm Dec} = +35\degr 39\arcmin 15\arcsec$,
and $z=0.00362$.  The host galaxy morphology is S0 and it exhibits a
faint spiral structure.  The distance to NGC\,5273 has been determined
from surface brightness fluctuations of the host galaxy
(\citealt{tonry01}, recalibrated by \citealt{tully08}) and is found to
be $16.5\pm1.6$\,Mpc.

\subsection{Spectroscopy}

Spectroscopic monitoring of NGC\,5273 was carried out at the Apache
Point Observatory 3.5\,m telescope from 2014 May 14 -- July 1 (UT
dates here and throughout).  The monitoring program was scheduled for
the first hour of almost every night during this time period,
resulting in observations being taken during evening twilight.  The
Dual Imaging Spectrograph (DIS) was employed with the low-resolution
(B400 and R300) gratings and the 5\arcsec\ slit rotated to a position
angle of 0\degr\ (oriented N$-$S).  With central wavelengths of
4398\,\AA\ and 7493\,\AA, the spectra span the entire optical bandpass
from the atmospheric cutoff to 1\,micron with a nominal dispersion of
1.8\,\AA/pix and 2.3\,\AA/pix, respectively.  A single spectrum with a
typical exposure time of 420\,s was obtained each night the weather
permitted, except for a span of 4 nights in June when the target was
above the 85\degr\ altitude limit of the telescope during the time
constraints of our program.  The spectrophotometric standard star,
Feige\,66, was also observed throughout this campaign, as were two
additional AGNs (we will describe their results elsewhere).

Spectra were reduced and flux calibrated following standard
procedures.  An extraction width of 12 pixels was used, corresponding
to an angular width of 5\arcsec\ and 4.8\arcsec\ for the blue and red
sides, respectively.  The regions for background subtraction
determination were set on either side of the spectrum between $75-95$
pixels ($\sim 30-40$\,arcsec) from the nucleus.

Spectroscopic monitoring campaigns generally do not enjoy the luxury
of photometric conditions every night, so the rough flux calibration
obtained with a spectrophometric standard star is not accurate enough
to measure the few-percent variations we expect during a reverberation
campaign.  A final, internal calibration of the spectra is required.
We follow the typical procedure, which is to intercalibrate the
spectra with the scaling method of \citet{vangroningen92} by assuming
a constant flux in the narrow emission lines.  A reference spectrum is
built from the user-determined best spectra, and the differences
between each individual spectrum and the reference are minimized in
the $\lambda\lambda$\,4959,5007 region.  This method accounts for
small differences in wavelength calibration, spectral resolution, and
flux calibration from night to night and has been shown to result in
relative spectrophotometry that is accurate to 2\%
\citep{peterson98a}. While variability of AGN narrow emission lines
has been detected \citep{peterson13}, the timescale of variation was
much longer (a few years) than the timescales probed by typical
reverberation-mapping campaigns such as the one we present here.  For
the red spectra, we instead used the [\ion{S}{2}] doublet for the
internal calibration.  The weakness of these emission lines relative
to H$\alpha$ resulted in a less accurate calibration for the spectra
and noisier light curves, but we were still able to measure a mean
time delay, as we discuss in $\S 3$.

\begin{figure}
\plotone{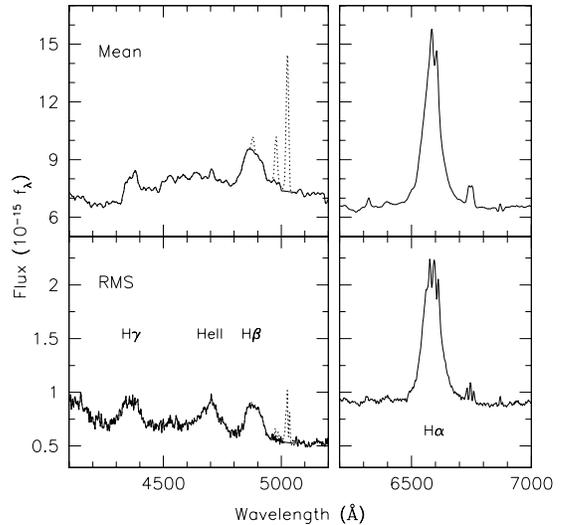}
\caption{Mean and root-mean-square (RMS) spectra of NGC\,5273 created
  from the $30$ individual spectra collected throughout the monitoring
  campaign. The solid line shows the narrow-line subtracted spectra,
  and the dotted line shows the spectra with the narrow lines
  included.  The RMS spectrum displays the variable components of the
  spectra.  The broad-emission lines are clearly visible, including
  \ion{He}{2} which is quite difficult to see in the mean spectrum.
  The host-galaxy features and narrow emission lines, which we expect
  to be constant on these timescales, are generally absent in the RMS
  spectrum except for some residual noise.}
\label{fig:spectra}
\end{figure}

Figure~\ref{fig:spectra} shows the final mean and root-mean-square
(RMS) of the 30 calibrated spectra collected over the 7 week period of
our monitoring program.  The Balmer lines H$\alpha$, H$\beta$, and
H$\gamma$, as well as \ion{He}{2} $\lambda 4686$ are clearly visible
in the RMS spectrum, evidencing their large variability during the
monitoring campaign.  Based on the nights where our campaign enjoyed
the best weather and observing conditions, we measured a mean
integrated flux of $f = 93.2 \pm 0.6 \times
10^{-15}$\,ergs\,s$^{-1}$\,cm$^{-2}$ for the [\ion{O}{3}] $\lambda
5007$ emission line, which agrees well with the value of $89.9 \times
10^{-15}$\,ergs\,s$^{-1}$\,cm$^{-2}$ determined by the Sloan Digital
Sky Survey \citep{ahn14} and sets the absolute flux scale of our
spectra.

\subsection{Photometry}

\begin{figure*}[ht!]
\plotone{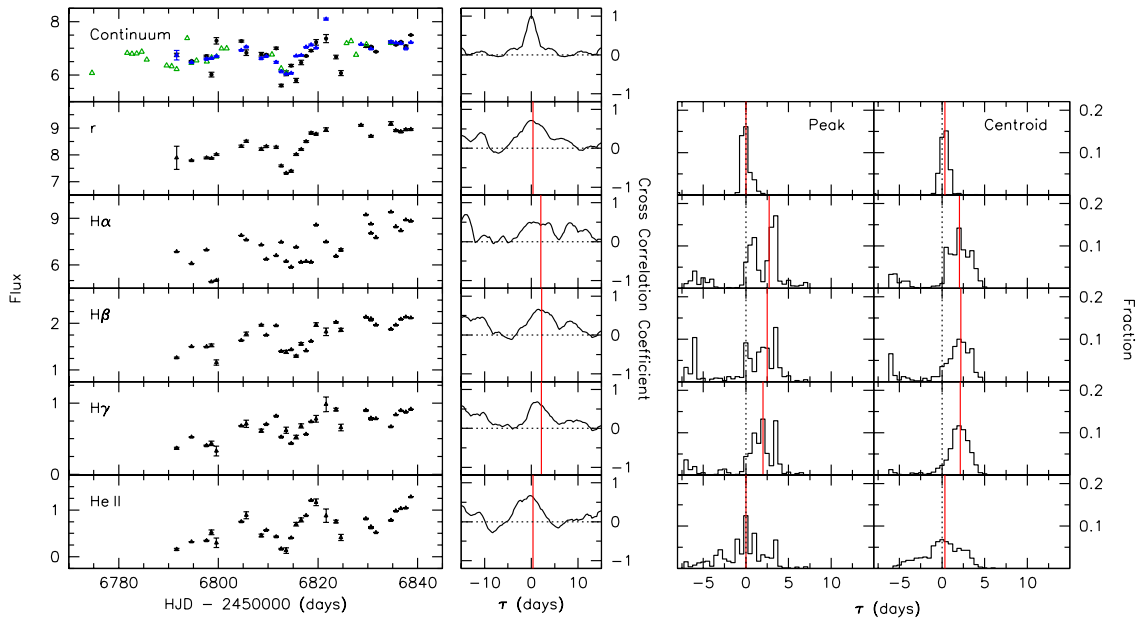}
\caption{({\it Left:}) Light curves and their corresponding
  cross-correlation functions relative to the continuum.  For the
  continuum light curve, this is the auto-correlation function.  The
  black points in the continuum box represent the $\lambda
  5100$\,\AA\ flux density, while the green triangles represent the
  HLCO $V-$band photometry and the blue points represent the APO
  $g-$band photometry.  The red vertical lines in the
  cross-correlation plots show the value of $\tau_{\rm cent}$ measured
  for each light curve from the distributions built up through the
  Monte Carlo techniques described in the text. ({\it Right:})
  Cross-correlation peak distributions and cross-correlation centroid
  distributions from which the mean time delays and their
  uncertainties are determined.}
\label{fig:lc}
\end{figure*}

Photometric monitoring was carried out at two observatories through
the months of 2014 May and June.  We obtained test observations on
2014 April 26 and began nightly monitoring on 2014 May 4 at Georgia
State University's Hard Labor Creek\footnote{The origins of the name
  for Hard Labor Creek are unclear, but a Wikipedia entry links the
  etymology to either the slaves who labored in the nearby cotton
  fields, or to the Native Americans who had difficulties fording the
  creek.} Observatory in Hard Labor Creek State Park near Rutledge,
GA.  Even with the unpredictable spring weather, we were able to
obtain photometry on nearly half (25/59) of the nights.
Images were acquired with the 24-inch Miller Telescope and the Apogee
2048$\times$2048 detector through a Johnson $V$-band filter.  The
detector spans a field-of-view of 26\farcm3 $\times$ 26\farcm3 with a
pixel size of 0\farcs77.  Typical exposure times were 180\,s, and
three dithered images were generally taken at the position of
NGC\,5273.  Images were reduced following standard procedures in {\tt
  IRAF.}\footnote{IRAF is distributed by the National Optical Astronomy
    Observatory, which is operated by the Association of Universities
    for Research in Astronomy (AURA) under cooperative agreement with
    the National Science Foundation.}

$V-$band light curves were derived from the individual images by
registering all images to a common alignment using {\tt Sexterp}
\citep{siverd12} and applying the image subtraction software package
{\tt ISIS }\citep{alard98,alard00}.  {\tt ISIS} builds a
reference frame from the images (defined by the user) with the best
seeing and lowest background levels.  It then uses a
spatially-variable kernel to convolve the reference frame to match the
point spread function (PSF) of each individual image.  Subtracting the
convolved reference frame leaves behind residuals that show any
regions of variability.  The light curve was measured from the
subtracted images, and therefore does not include any constant flux
components such as the contribution from the host galaxy starlight.

Photometry was also obtained at the Apache Point Observatory 3.5\,m
telescope in New Mexico.  A 30\,s image was typically obtained
immediately following readout of the spectra.  The two arms of the DIS
instrument allowed two images --- Gunn-Thuan $g$ and $r$ bands --- to
be obtained simultaneously with the blue and red cameras.  On a few
nights, images were not obtained after the spectra.  These nights were
typically partly cloudy, leading to a delay in pointing and focusing
the telescope and shortening the on-sky time for our program.  Images
were reduced in {\tt IRAF} following standard procedures.  The narrow
field of view ($\sim 4\farcm1 \times 6\farcm9$) and small number of
field stars in each DIS image precluded the use of image subtraction
techniques.  Instead, we carried out aperture photometry of NGC\,5273
and a comparison field star 132\arcsec\ to the southwest
(SDSS\,J134202.91+353721.6).  We employed circular apertures with
radii of 2\farcs1 ($g$) and 2\farcs0 ($r$), and sky annuli of
$4\farcs6-5\farcs9$ ($g$) and $4\farcs4-5\farcs6$ ($r$).

\section{Light Curve Analysis}

Emission-line light curves were determined from the final, scaled
spectra by fitting a local, linear continuum under each emission line
and integrating the line flux above this continuum.  We do not follow
the procedure of fitting the entire spectrum with model components
because the results are very sensitive to the exact parameters that
are included and how they are modeled \citep{denney09a}.  The light
curve for the continuum flux density at 5100\,\AA $\times (1+z)$ was
also measured.  Table~\ref{tab:lc} gives our tabulated spectroscopic
and photometric light curves.

The $V-$band filter ($\lambda_c = 5483$\,\AA\ and $\Delta
\lambda=827$\,\AA) does not include a significant contribution from
any of the broad emission lines in the spectrum of NGC\,5273, so we
combined the $V-$band photometry with the 5100\,\AA\ continuum flux
densities.  We identified pairs of points in the two light curves that
were contemporaneous within 0.75\,days. A linear relationship was then
fit to the pairs of points to determine the multiplicative and
additive factors necessary to bring the $V-$band fluxes into agreement
with the 5100\,\AA\ fluxes, taking into account the differences in
galaxy background light and bandpass.

We then compared the merged continuum light curve with the Gunn-Thuan
$g-$ and $r-$band photometry from APO.  We detected no significant
time delay between the continuum and the $g-$band, so we merged those
light curves together in the same way.  However, we did detect a small
time delay between the continuum light curve and the $r-$band
photometry (see Figure~\ref{fig:lc} and Table~\ref{tab:lagwidth}, so
we did not merge the $r-$band with the continuum light curve.  The
small delay determined for the $r-$band is likely due to the strong
contribution to the bandpass from the broad H$\alpha$ emission line
($\sim 90$\% of the $r-$band flux,
  cf.\ Table~\ref{tab:lcstats}).  A similar effect would be expected
from H$\beta$ for the $g-$band, but the smaller equivalent width of
H$\beta$ relative to the total bandpass ($\sim 40$\% of the
  $g-$band flux) is likely the reason we did not detect any time
delay for $g$.

Figure~\ref{fig:lc} shows the light curves for the continuum
(including the $V-$band and $g-$band photometry), the $r-$band
photometry, and for the optical recombination lines H$\alpha$,
H$\beta$, H$\gamma$, and \ion{He}{2}.  NGC\,5273 exhibited fairly
strong variability over the course of the monitoring campaign.
Table~\ref{tab:lcstats} gives the variability statistics for the final
light curves displayed in Figure~\ref{fig:lc}.  Column (1) lists the
spectral feature and column (2) gives the number of measurements in
the light curve.  Columns (3) and (4) list the average and median time
separation between measurements, respectively.  Column (5) gives the
mean flux and standard deviation of the light curve, and column (6)
lists the mean fractional error (based on the comparison of
observations that are closely spaced in time).  Column (7) lists the
excess variance, computed as:
\begin{equation}
F_{\rm var} = \frac{\sqrt{\sigma^2 - \delta^2}}{\langle F \rangle}
\end{equation}
\noindent where $\sigma^2$ is the variance of the fluxes, $\delta^2$
is their mean-square uncertainty, and $\langle F \rangle$ is the mean
flux. And column (8) is the ratio of the maximum to the minimum flux
in the light curve.  In general, the true level of variability was
somewhat higher than these measurements describe if taken at face
value.  The continuum light curve includes the host-galaxy starlight
contribution, as does the $r-$band light curve.  The broad
emission-line light curves include the contribution from the narrow
emission lines.  These non-variable components serve to dampen the
variations we observe in the light cuves.

To determine the average time lag of the emission lines relative to
the continuum, we first cross-correlated the light curves using the
interpolation cross-correlation method \citep{gaskell86,gaskell87}
with the modifications of \citet{white94}.  The method determines the
cross-correlation function (CCF) twice: first, by interpolating the
continuum light curve, and then by interpolating the emission-line
light curve.  The final CCF is the average of the two.  We
characterize the results of the CCF by recording the maximum value of
the CCF ($r_{\rm max}$ ), the time delay of the CCF maximum
($\tau_{\rm peak}$) and the centroid of the points about the peak
($\tau_{\rm cent}$) above a threshold value of $0.8 r_{\rm max}$.
CCFs for each light curve relative to the continuum are displayed in
Figure~\ref{fig:lc}.

To quantify the uncertainties on the time lag measurements, we employ
the Monte Carlo ``flux randomization/random subset sampling'' method
of \citet{peterson98b,peterson04}.  From the $N$ available data
points, a selection of $N$ points is chosen without regard to whether
a datum has been previously selected.  The uncertainty on a point that
is sampled $1 \leq n \leq N$ times is scaled by a factor of $n^{1/2}$
and the typical number of points that is not sampled in any specific
realization is $\sim 1/e$.  This is the ``random subset sampling'',
and it quantifies the uncertainty that arises from the inclusion of
any specific data point in the light curve.  The flux values in the
selected subset are then modified by a Gaussian deviation of the flux
uncertainty.  This is the ``flux randomization'' and it accounts for
the measurement uncertainties.  The final sampled and modified light
curves are then cross-correlated and the CCF measurements are
recorded.  The process is repeated many times ($N = 1000$) and a
distribution of CCF measurements are built up (see the right-hand
  panels in Figure~\ref{fig:lc}).  We take the means of the
cross-correlation centroid distribution and the cross-correlation peak
distribution as $\tau_{\rm cent}$ and $\tau_{\rm peak}$, respectively.
The uncertainties on $\tau_{\rm cent}$ and $\tau_{\rm peak}$ are
determined so that 15.87\% of the realizations fall above and 15.87\%
fall below the range of uncertainties, corresponding to $\pm 1 \sigma$
for a Gaussian distribution.

Table~\ref{tab:lagwidth} lists the measured time lags and their
uncertainties for the emission lines and for the $r-$band relative to
the continuum light curve.  We also investigated the time lags with
the software package {\tt JAVELIN}, previously known as {\tt SPEAR}
\citep{zu11}.  {\tt JAVELIN} uses a damped-random walk to fit the
continuum light curve, and then determines the best model for the
reprocessed (shifted and smoothed) emission-line light curves by
maximizing the likelihood of the model.  Uncertainties in the time
delay are determined through a Bayesian Markov Chain Monte Carlo
method.

\begin{figure}
\plotone{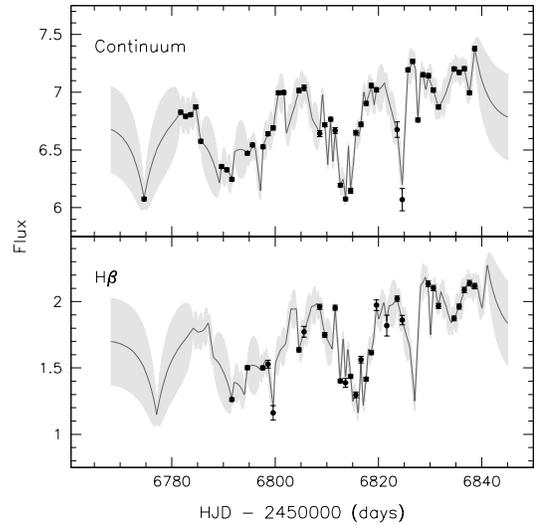}
\caption{Continuum and H$\beta$ light curves (data points) with the
  mean {\tt JAVELIN} model (solid lines) and uncertainties (gray
  shaded regions).  The uncertainties on the mean model are derived
  from the standard deviation of the individual realizations.}
\label{fig:lc.jav}
\end{figure}

Unfortunately, we were unable to obtain reasonable constraints on the
smoothing and delay parameters when simultaneously modeling multiple
emission-line light curves with {\tt JAVELIN}.  Instead, we ran
multiple models and each time we focused on modeling the continuum and
a single emission line.  Figure~\ref{fig:lc.jav} shows the modeling
results obtained for the H$\beta$ emission-line light curve.  The time
lags we obtained with {\tt JAVELIN} were generally consistent with
those determined through cross-correlation methods (see
Table~\ref{tab:lagwidth}), but the difficulties we encoutered leads us
to focus on the time lags obtained through cross-correlation methods
throughout the remainder of this manuscript.

\begin{figure*}[ht!]
\plotone{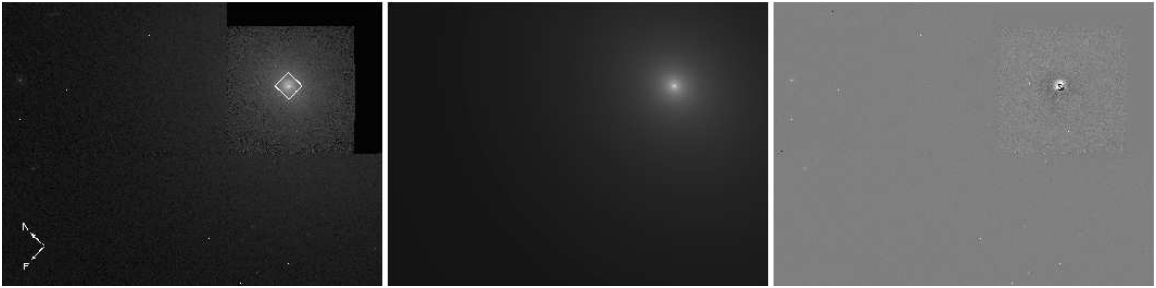}
\caption{{\it HST} WFPC2 image of NGC\,5273 with the 5\arcsec $\times$
  5\arcsec\ spectroscopic monitoring slit geometry superposed ({\it
    left}).  The best-fit model from two-dimensional surface
  brightness modeling with Galfit ({\it middle}) and the residuals
  after subtraction of the model from the image ({\it right}) are also
  shown.  Regions with no data (such as the area outside the PC
  detector) were masked during the fitting process.}
\label{fig:galfit}
\end{figure*}

\begin{figure}[hb!]
\plotone{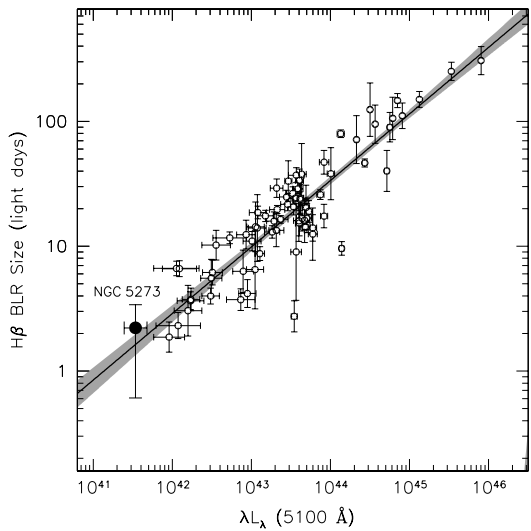}
\caption{The H$\beta$ \rl\ relationship for the reverberation sample
  of AGNs.  Open circles show the measured time delays and
  luminosities as tabulated in \citet{bentz13}, and the power law and
  gray shaded region are the best fit and 1$\sigma$ uncertainty for
  the \citet{bentz13} compilation.  The solid point shows our
  measurement of the H$\beta$ time delay and starlight corrected
  luminosity for NGC\,5273 from this work.}
\label{fig:rl}
\end{figure}

Finally, we can compare the time delay measured for H$\beta$ with that
expected from the relationship between emission line time delay and
AGN luminosity (the \rl\ relationship) determined for the full
reverberation sample of AGNs.  The 5100\,\AA\ luminosity of NGC\,5273
as determined from the mean spectrum contains a large contribution
from host-galaxy starlight.  Fortuitously, the {\it Hubble Space
  Telescope} archive contains WFPC2 imaging of NGC\,5273 through the
F547M filter, which allowed us to determine the starlight correction
and deduce the AGN luminosity.  Following the methods of
\citet{bentz13}, we modeled the surface brightness distribution of the
drizzled and combined WFPC2 image in two dimensions with Galfit
\citep{peng02,peng10}.  The galaxy was well fit with an exponential
disk and a \sersic\ bulge with index $n=3.3$.  The modeling allowed us
to accurately subtract the AGN contribution and create a
``PSF-free'' image of NGC\,5273, from which we determined the
starlight flux through the monitoring aperture of $(4.80 \pm 0.48)
\times 10^{-15}$\,ergs\,s$^{-1}$\,cm$^{-1}$\,\AA\ at $5100 \times
(1+z)$\,\AA.  The mean flux at 5100\,\AA\ (as shown in
Table~\ref{tab:lcstats}) was corrected for the starlight contribution,
giving an AGN luminosity of  $\log L_{\rm AGN} / {\rm ergs~
    s^{-1}} = 41.534 \pm 0.144$.  Figure~\ref{fig:rl} shows our
H$\beta$ time delay measurement and starlight corrected luminosity for
NGC\,5273, which agrees well with the expectations from the
\rl\ relationship.

\section{Line Width Measurements}

The width of a broad emission line is a measure of the line-of-sight
velocity of the gas in the broad line region.  Narrow emission lines
are emitted from gas that does not participate in the same bulk motion
as the broad line region gas, so we report the width of only the broad
components of the emission lines.

Figure~\ref{fig:spectra} shows the mean and RMS spectra for NGC\,5273
(dotted lines) as well as the narrow-line subtracted mean and RMS
spectra (solid lines).  For the narrow line subtraction, we used the
[\ion{O}{3}] $\lambda 5007$ emission line as a template for the
$\lambda 4959$ and H$\beta$ narrow lines. The ratio of
[\ion{O}{3}]~$\lambda 4959$/[\ion{O}{3}]~$\lambda 5007$ was set at
0.34 \citep{storey00}, and we derived a ratio of
H$\beta$/[\ion{O}{3}]~$\lambda 5007 = 0.10$.  Due to the lack of a
suitable narrow emission line in the red spectra for use as a
template, we did not attempt any narrow line subtraction with
H$\alpha$.

The widths of the broad emission lines were measured in the
narrow-line subtracted mean and RMS spectra and are reported as two
separate measures: the full width at half-maximum (FWHM) flux, and the
line dispersion, $\sigma_{\rm line}$, which is the second moment of
the emission-line profile \citep{peterson04}. The uncertainties in the
line widths were set using a Monte Carlo random subset sampling
method. In this case, from a set of $N$ spectra, a random subset of
$N$ spectra were selected without regard to whether a spectrum had
previously been chosen, and mean and rms spectra were created from the
chosen subset. The FWHM and the $\sigma_{\rm line}$ were measured and
recorded for each realization, and distributions of line-width
measurements were built up over 1000 realizations. The mean and
standard deviation of each distribution are taken to be the line width
and uncertainty, respectively.  This method quantifies the weight that
any individual spectrum has on the final line width measurements.
Additionally we also quantify the uncertainty from the exact placement
of the continuum region on either side. For each line, we define a
maximum continuum window (typically $30-50$\,\AA\ wide) on either side
of an emission line.  For each realization, a subset of each continuum
window of at least 7 pixels ($\sim 14$\,\AA) is randomly selected,
from which the local linear continuum is fit.  This additional step
has little effect on the RMS linewidths, because their uncertainties
are dominated by the noise in the spectra, but has a small effect 
  that increases the uncertainties we measure for the linewidths
derived from the mean spectra \citep{bentz09c}.

\begin{figure*}[ht!]
\plottwo{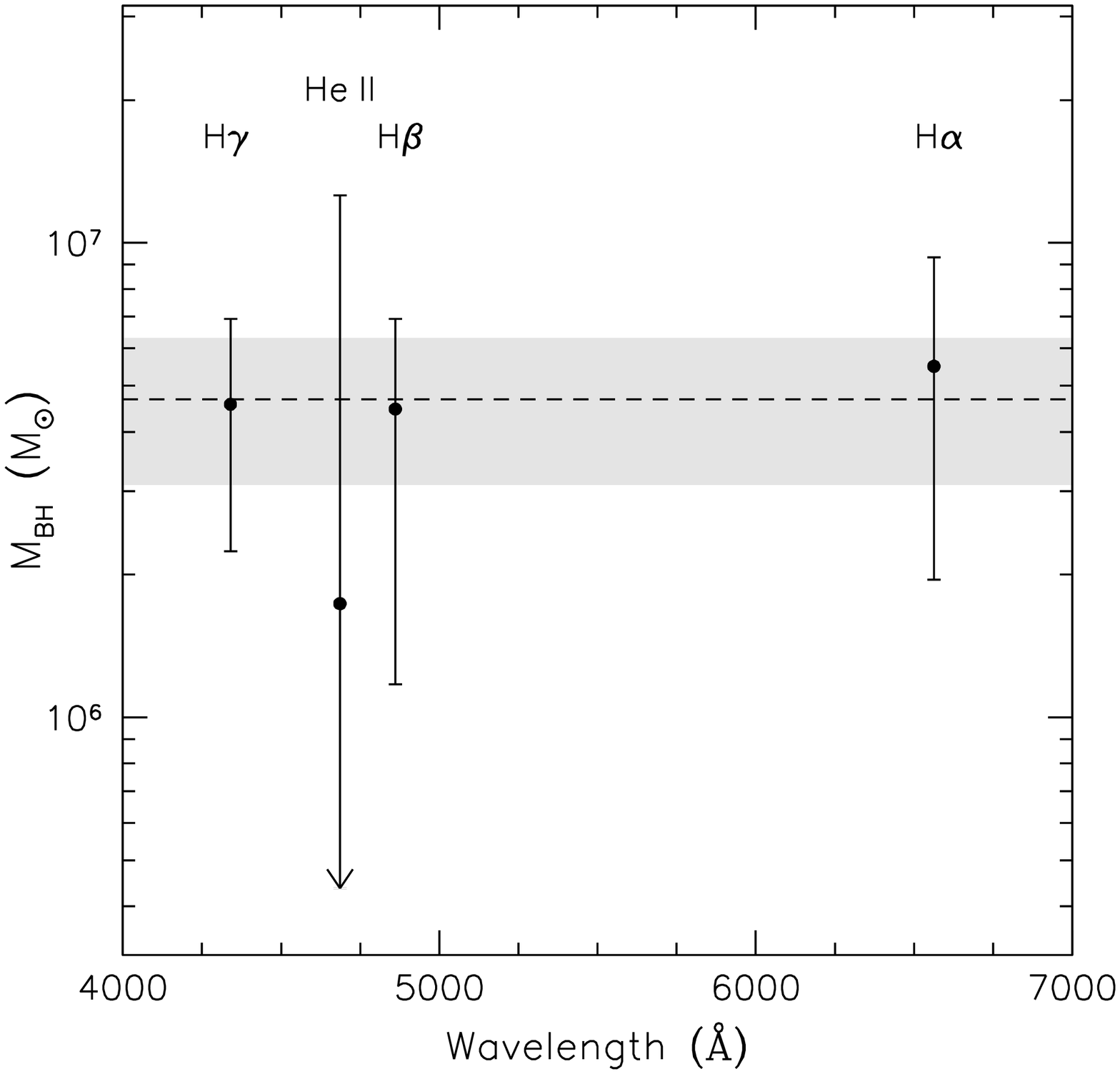}{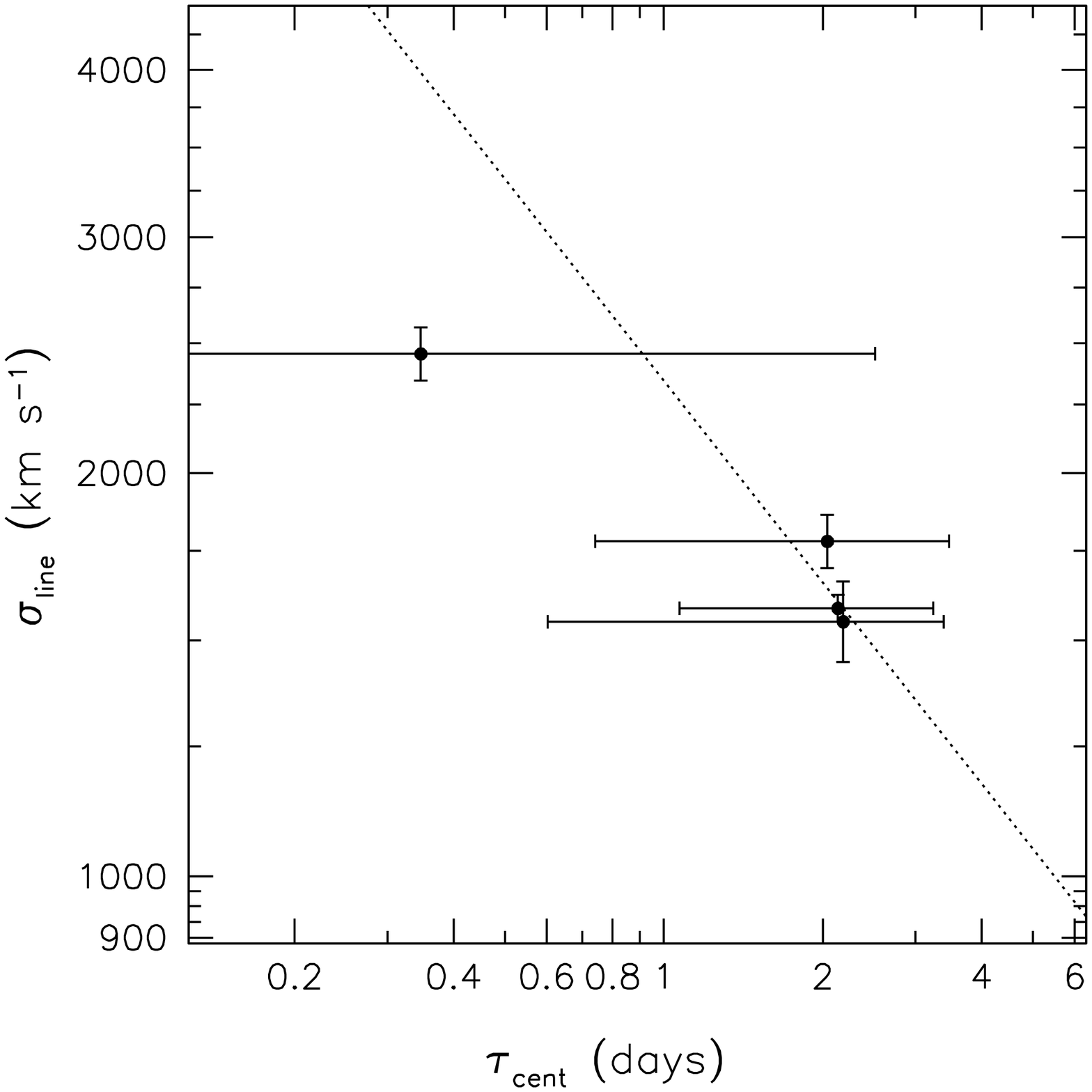}
\caption{{\it Left:} Black hole masses derived from the emission line
  time delays and velocity widths.  Each individual mass constraint is
  plotted at the central wavelength of the emission line from which it
  was derived. The dashed horizontal line and the shaded grey region
  show the weighted mean of the black hole mass and the $1\sigma$
  uncertainty, respectively.  {\it Right:} Velocity widths of the
  emission lines compared to their measured time delays.  The dotted
  line is not a fit to the points, but is a power law with slope of
  $-0.5$ and intercept assuming the weighted mean black hole mass we
  have determined for NGC\,5273.  It shows the expected behavior for
  gas with motions dominated by the gravity of the black hole.}
\label{fig:mbh}
\end{figure*}

Emission-line widths were corrected for the dispersion of the
spectrograph following \citet{peterson04}.  Specifically, the observed
line width $\Delta \lambda_{\rm obs}$ can be described as a
combination of the intrinsic line width, $\Delta \lambda_{\rm true}$,
and the spectrograph dispersion, $\Delta \lambda_{\rm disp}$, such
that
\begin{equation}
\Delta \lambda_{\rm obs}^2 \approx \Delta \lambda_{\rm true}^2 + \Delta \lambda_{\rm disp}^2.
\end{equation}
\noindent We determined $\Delta \lambda_{\rm true}$ by taking our
measurement of the FWHM of [\ion{O}{3}] $\lambda 5007$ as $\Delta
\lambda_{\rm obs} = 14.26$\,\AA\ and the FWHM measured by
\citet{whittle92} through a small slit and with a high resolution as
$\Delta \lambda_{\rm true} = 130$\,km\,s$^{-1}$.  We deduced a final
spectral resolution of $\Delta \lambda_{\rm disp} = 14.1$\,\AA, which
we used to correct our line width measurements.  The final rest-frame,
resolution-corrected line width measurements determined from the mean
and the RMS spectra are listed in Table~\ref{tab:lagwidth}.

\section{Black Hole Mass}

The black hole mass is generally determined from reverberation-mapping
measurements as:
\begin{equation}
M_{\rm BH} = f \frac{c \tau V^2}{G}
\end{equation}
where $\tau$ is the time delay for a specific emission line relative
to continuum variations, and $V$ is the line-of-sight velocity width
of the emission line, with $c$ and $G$ being the speed of light and
gravitational constants, respectively.  The emission-line time delay
is therefore a measure of the responsivity-weighted average radius of
the broad-line region for the emission of a particular species (e.g.,
H$\beta$).  Only gas that is optically thick to ionizing radiation
will be included in this measure, because optically thin gas will not
respond to changes in the ionizing flux and will not reverberate.


The factor $f$ includes the details of the inclination of the system
to our line of sight and the exact geometrical arrangement of the
responding gas, as well as its kinematics.  These details are
generally unknown for any particular object because the gas only
extends across angular scales of milliarcseconds or less from our
vantage point.  Current practice is to apply a population-averaged
value, $\langle f \rangle$, to reverberation masses to get the overall
mass scale correct for the sample as a whole.  The value of $\langle f
\rangle$ is determined from a comparison of the \msigma\ relationship
for dynamical black hole masses and the \msigma\ relationship for AGNs
with reverberation measurements.  The overall multiplicative factor
that must be applied to the AGN masses to bring the two relationships
into agreement is taken to be $\langle f \rangle$.  This value has
varied in the literature from 5.5 \citep{onken04} to 2.8
\citep{graham11}, depending on which objects are included and the
specifics of the measurements.  We adopt the value determined by
\citet{grier13} of $\langle f \rangle = 4.3 \pm 1.1$ as it uses the
most up-to-date set of measurements, and includes four high-luminosity
AGNs with large \mbh\ that better anchor the high-mass end of the AGN
\msigma\ relationship.

While the use of the \msigma\ relationship to set the absolute mass
scale for reverberation masses has been fairly standard for the past
10 years, we note that there are several potential problems with this
method.  In particular, several studies have uncovered large scatter
and morphological biases in the \msigma\ relationship for quiescent
galaxies (e.g., \citealt{hu08,graham11,kormendy11}).  There are
several lines of evidence that demonstrate, however, that adopting a
mean scaling factor from comparison of \msigma\ relationships does put
reverberation masses in the correct vicinity.  \citet{bentz09a}
compared the \ml\ relationship for reverberation-mapped AGNs and
quiescent galaxies and found that the two relationships were
consistent when an average scaling factor determined from the
\msigma\ relationship was adopted.  \citet{pancoast14} directly
determined AGN black hole masses through dynamical modeling of
reverberation-mapping spectra and found an average $\langle f \rangle
= 4.8$ for the five AGNs they examined.  They were also able to
constrain the AGN inclinations, one of the largest expected
contributions to the value of $\langle f \rangle$, and found
inclinations of $10-50$\degr\ to our line of sight.  This agrees well
with the inclination of $\sim 29$\degr\ implied by a scaling factor of
$\langle f \rangle = 4.3$, if other effects are neglected.
Additionally, \citet{fischer13} analyzed the spatially-resolved
biconical narrow line regions of several nearby AGNs with
three-dimensional geometric models and found that the Seyfert 1s in
their sample had inclinations to our line of sight of
$12-49$\degr\ with an average inclination of $\sim 24$\degr.  All of
these independent studies suggest that adopting $\langle f \rangle =
4.3$ for reverberation masses will result in a sample of unbiased
masses, although the mass of any particular AGN is likely only
accurate to a factor of 2-3.

Each emission line in our analysis provides an independent measurement
of the black hole mass in NGC\,5273.  We can investigate the
reliability of a reverberation-based black hole mass determination by
comparing the results for all the emission lines. Figure~\ref{fig:mbh}
(left) shows the black hole mass derived from the time delay and line
width of each of the optical recombination lines we were able to
probe: H$\alpha$, H$\beta$, H$\gamma$, and \ion{He}{2}.  The right
panel of Figure~\ref{fig:mbh} shows the emission line width versus the
time delay for each line.  The dotted line is a power law with slope
of $-0.5$ which is the expected relationship if the gravity of the
black hole dominates the dynamics of the gas we are probing.  The
results are consistent with the expectation, as has been seen for
other reverberation studies (e.g., \citealt{peterson04,bentz09c}).

Combining our time lags ($\tau_{\rm cent}$) and line widths
($\sigma_{\rm line,rms}$) for each emission line and scaling by
$\langle f \rangle$, we determine a final weighted mean of the black
hole mass in NGC\,5273 of $M_{\rm BH} = (4.7 \pm 1.6) \times
  10^6$\,M$_{\odot}$.

\section{Discussion}

The sphere of influence of the black hole is generally defined as:
\begin{equation}
r_{\rm h} = \frac{GM_{\rm BH}}{\sigma_{\star}^2}
\end{equation}
and it is a useful metric for comparing with the spatial resolution of
stellar dynamical observations, to determine whether or not a reliable
black hole mass is likely to be obtained from dynamical modeling.  

\begin{figure}[ht!]
\plotone{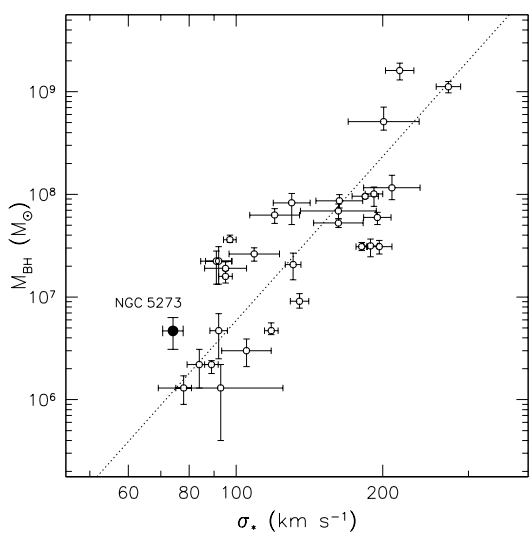}
\caption{The \msigma\ relationship for AGNs from \citet{grier13} with
  the best-fit relationship found for quiescent galaxies with
  dynamical masses \citep{woo13}.  The solid point is our measurement
  for NGC\,5273.  While slightly above the relationship, NGC\,5273 is
  within the scatter exhibited by the other AGNs.}
\label{fig:msig}
\end{figure}

NGC\,5273 was included in the ATLAS$^{3D}$ \citep{cappellari11} sample
of early-type galaxies.  \citet{cappellari13} derive a bulge stellar
velocity dispersion of $\sigma (R_e/8) = 74.1 \pm 3.7$\,km\,s$^{-1}$
from integral-field spectroscopy of the galaxy.  Combined with the
black hole mass we have derived here, we estimate $r_{\rm h} =
0.05$\arcsec. This estimate depends rather sensitively on the
specific value of $\langle f \rangle$ that is adopted.
Figure~\ref{fig:msig} shows where NGC\,5273 lies on the most recent
\msigma\ relationship for AGNs, as determined by \citet{grier13}.
NGC\,5273 sits slightly above the relationship, however it is within
the scatter, lending credence to the value of \mbh\ determined here
with $\langle f \rangle = 4.3$.

Current ground-based large-aperture telescopes are capable of
obtaining spatial resolutions of $\sim 0.08$\arcsec\ in the
near-infrared with adaptive optics and integral-field units. The James
Webb Space Telescope (JWST) will be able to achieve a similar
resolution, but will have the added advantage of being located above
the Earth's atmosphere, allowing for a much lower background and a
more stable PSF with a significantly higher Strehl ratio.
\citet{gultekin09} show that a strict insistence on resolving $r_h$ is
not necessary for determination of an accurate stellar dynamical mass,
which seems to be evidenced by the quality of the dynamical mass
obtained for NGC\,3227 (where $r_h$ is not resolved;
\citealt{davies06}).  The early-type host galaxy of NGC\,5273 makes it
the ideal candidate for stellar dynamical modeling in order to
directly test the masses obtained from reverberation mapping and those
obtained from stellar dynamical modeling.

It is interesting that NGC\,5273 sits above the relationship, as it is
hosted by an early-type galaxy.  The AGNs sitting slightly below the
relationship at similar black hole masses all live in barred spiral
galaxies, and \citet{hu08} and \citet{graham11} have found that barred
or pseudobulge galaxies have black hole masses that lie preferentially
below the \msigma\ relationship for early-type galaxies.  It is
currently unclear, however, whether this is because the black holes
are undermassive or because the bar dynamics are artificially
broadening the bulge stellar absorption line signatures in the
spectra.  The majority of the bulge stellar velocity dispersions in
the AGN sample with reverberation masses are determined from fitting
the stellar absorption lines in a single long slit spectrum, so it is
likely that there is contamination from the bar dynamics in many of
the stellar velocity dispersion measurements for the AGNs.
\citet{grier13} attempted to quantify such a bias, but were unable to
detect it among the current sample.  On the other hand,
\citet{onken14} found that using an axisymmetric dynamical modeling
code to determine a stellar dynamical black hole mass could lead to a
biased black hole mass measurement in barred galaxies, although in the
case of NGC\,4151, the object in their study, the bias led to an {\it
  overestimate} of \mbh, not an underestimate.  Clearly, additional
study is necessary to determine the exact effects caused by galaxy
bars on black hole mass determinations, and NGC\,5273 is an important
addition to the reverberation sample given its proximity and the
unbarred early-type morphology of its host galaxy.


\section{Summary}

We have carried out a spectroscopic and photometric monitoring
campaign of the AGN in the nearby Seyfert galaxy NGC\,5273.  From the
time delays measured between the broad optical recombination lines and
the continuum flux, we determine a black hole mass of $M_{\rm BH} =
(4.7 \pm 1.6) \times 10^6$\,M$_{\odot}$.  Combined with the bulge
stellar velocity dispersion, we estimate that the black hole sphere of
influence for NGC\,5273 should be just at the limit of the resolution
achievable with current ground-based instrumentation and with the
integral field capabilities of JWST.  NGC\,5273 is the newest addition
to a very short list of AGNs with reverberation-based masses and black
hole spheres of influence capable of being probed with current
technology.  It is also the only one of these few AGNs in an
early-type unbarred galaxy, which makes NGC\,5273 an obvious candidate
for stellar dynamical modeling for a direct comparison of the black
hole masses determined from different techniques.

\acknowledgements 

We thank the anonymous referee for comments and suggestions that
improved the presentation of this manuscript.  We thank Chris Onken,
Kelly Denney, Kate Grier, Gisella de Rosa, and Ying Zu for helpful
feedback and discussions.  This research is based on observations
obtained with the Apache Point Observatory 3.5-meter telescope, which
is owned and operated by the Astrophysical Research Consortium.  We
heartily thank the staff at APO --- especially telescope operators
Alaina, Alysha, Jack, and Russet --- for all their help with this
program.  It was a pleasure to interact with y'all every night for two
months!  MCB gratefully acknowledges support from the NSF through
CAREER grant AST-1253702. This research has made use of the NASA/IPAC
Extragalactic Database (NED) which is operated by the Jet Propulsion
Laboratory, California Institute of Technology, under contract with
the National Aeronautics and Space Administration and the SIMBAD
database, operated at CDS, Strasbourg, France.

\bibliographystyle{apj} 

\clearpage

\begin{landscape}
\begin{deluxetable}{lcccccccccccc}
\tablecolumns{11}
\tablecaption{Tabulated Light Curves}
\tablehead{
\colhead{HJD} &
\colhead{$\lambda 5100$\,\AA\tablenotemark{a}} &
\colhead{H$\alpha$\tablenotemark{b}} &
\colhead{H$\beta$\tablenotemark{b}} &
\colhead{H$\gamma$\tablenotemark{b}} &
\colhead{\ion{He}{2}\tablenotemark{b}} &
\colhead{HJD} &
\colhead{$g$} &
\colhead{$r$} &
\colhead{HJD} &
\colhead{$V$\tablenotemark{c}} \\
\colhead{(days)} &
\colhead{} &
\colhead{} &
\colhead{} &
\colhead{} &
\colhead{} &
\colhead{(days)} &
\colhead{(mag)} &
\colhead{(mag)} &
\colhead{(days)} &
\colhead{}
}
\startdata
6791.6275  & $6.754 \pm	0.064$ & $6.857	\pm 0.015$ & $1.262 \pm	0.013$ & $0.367	\pm 0.014$ & $0.157 \pm	0.022$ & 6791.63 & $14.771 \pm	0.047$ & $14.169 \pm 0.058$ & 6774.6377 & $6.075 \pm 0.010$ \\ 
6794.6247  & $6.523 \pm	0.019$ & $6.100	\pm 0.001$ & $1.502 \pm	0.001$ & $0.524	\pm 0.001$ & $0.319 \pm	0.002$ & 6794.63 & $14.846 \pm	0.004$ & $14.183 \pm 0.003$ & 6781.7165 & $6.827 \pm 0.012$ \\
6797.6143  & $6.711 \pm	0.036$ & $6.974	\pm 0.005$ & $1.501 \pm	0.005$ & $0.404	\pm 0.006$ & $0.345 \pm	0.009$ & 6797.62 & $14.810 \pm	0.007$ & $14.169 \pm 0.004$ & 6782.6802 & $6.791 \pm 0.010$ \\
6798.6200  & $6.013 \pm	0.089$ & $4.928	\pm 0.022$ & $1.529 \pm	0.029$ & $0.435	\pm 0.032$ & $0.526 \pm	0.051$ & 6798.62 & $14.803 \pm	0.004$ & $14.172 \pm 0.004$ & 6783.6364 & $6.805 \pm 0.011$ \\
6799.6162  & $7.282 \pm	0.118$ & $5.019	\pm 0.021$ & $1.162 \pm	0.055$ & $0.329	\pm 0.068$ & $0.298 \pm	0.102$ & 6799.62 & $14.781 \pm	0.006$ & $14.152 \pm 0.004$ & 6784.6254 & $6.873 \pm 0.011$ \\
6804.6151  & $7.270 \pm	0.039$ & $7.899	\pm 0.003$ & $1.636 \pm	0.005$ & $0.679	\pm 0.006$ & $0.751 \pm	0.009$ & 6804.62 & $14.720 \pm	0.006$ & $14.111 \pm 0.005$ & 6785.6180 & $6.575 \pm 0.010$ \\
6805.6161  & $6.833 \pm	0.105$ & $7.619	\pm 0.022$ & $1.773 \pm	0.041$ & $0.710	\pm 0.052$ & $0.889 \pm	0.075$ & 6805.62 & $14.689 \pm	0.006$ & $14.087 \pm 0.005$ & 6789.6239 & $6.356 \pm 0.010$ \\
6808.6199  & $6.780 \pm	0.068$ & $7.296	\pm 0.007$ & $1.961 \pm	0.017$ & $0.612	\pm 0.021$ & $0.456 \pm	0.031$ & 6808.62 & $14.804 \pm	0.008$ & $14.125 \pm 0.006$ & 6790.6528 & $6.328 \pm 0.015$ \\
6809.6203  & $6.759 \pm	0.025$ & $6.371	\pm 0.013$ & $1.749 \pm	0.002$ & $0.702	\pm 0.003$ & $0.569 \pm	0.004$ & 6809.62 & $14.783 \pm	0.005$ & $14.112 \pm 0.004$ & 6791.6362 & $6.215 \pm 0.015$ \\
6811.6170  & $7.001 \pm	0.042$ & $6.605	\pm 0.003$ & $1.953 \pm	0.007$ & $0.819	\pm 0.009$ & $0.429 \pm	0.013$ & 6811.62 & $14.845 \pm	0.009$ & $14.116 \pm 0.005$ & 6793.7453 & $7.381 \pm 0.031$ \\
6812.6227  & $5.606 \pm	0.044$ & $7.466	\pm 0.002$ & $1.402 \pm	0.007$ & $0.524	\pm 0.008$ & $0.169 \pm	0.012$ & 6812.63 & $14.952 \pm	0.005$ & $14.211 \pm 0.004$ & 6794.6275 & $6.450 \pm 0.012$ \\
6813.6205  & $6.066 \pm	0.094$ & $6.232	\pm 0.020$ & $1.388 \pm	0.033$ & $0.619	\pm 0.042$ & $0.133 \pm	0.060$ & 6813.62 & $14.982 \pm	0.006$ & $14.251 \pm 0.004$ & 6795.6200 & $6.543 \pm 0.009$ \\
6814.6198  & $6.351 \pm	0.041$ & $5.865	\pm 0.003$ & $1.437 \pm	0.006$ & $0.433	\pm 0.007$ & $0.403 \pm	0.010$ & 6814.62 & $14.972 \pm	0.008$ & $14.240 \pm 0.006$ & 6797.7028 & $6.506 \pm 0.009$ \\
6815.6210  & $5.796 \pm	0.077$ & $7.147	\pm 0.030$ & $1.297 \pm	0.021$ & $0.524	\pm 0.026$ & $0.698 \pm	0.038$ & 6815.62 & $14.778 \pm	0.006$ & $14.152 \pm 0.004$ & 6798.6235 & $6.653 \pm 0.009$ \\
6816.6235  & $6.467 \pm	0.082$ & $6.180	\pm 0.015$ & $1.562 \pm	0.025$ & $0.680	\pm 0.031$ & $0.789 \pm	0.044$ & 6816.63 & $14.771 \pm	0.006$ & $14.127 \pm 0.004$ & 6799.6525 & $6.686 \pm 0.009$ \\
6817.6217  & $6.712 \pm	0.023$ & $6.229	\pm 0.007$ & $1.416 \pm	0.002$ & $0.562	\pm 0.002$ & $0.884 \pm	0.003$ & 6817.63 & $14.690 \pm	0.005$ & $14.088 \pm 0.004$ & 6800.6366 & $6.995 \pm 0.009$ \\
6818.6197  & $6.913 \pm	0.035$ & $6.187	\pm 0.014$ & $1.616 \pm	0.004$ & $0.738	\pm 0.005$ & $1.205 \pm	0.008$ & 6818.62 & $14.669 \pm	0.006$ & $14.048 \pm 0.005$ & 6801.6919 & $6.998 \pm 0.010$ \\
6819.6240  & $7.225 \pm	0.107$ & $8.574	\pm 0.025$ & $1.974 \pm	0.040$ & $0.779	\pm 0.047$ & $1.168 \pm	0.071$ & 6819.63 & $14.698 \pm	0.005$ & $14.053 \pm 0.004$ & 6810.7598 & $6.766 \pm 0.012$ \\
6821.6228  & $7.367 \pm	0.150$ & $7.487	\pm 0.026$ & $1.819 \pm	0.079$ & $0.982	\pm 0.102$ & $0.879 \pm	0.144$ & 6821.63 & $14.453 \pm	0.008$ & $14.033 \pm 0.008$ & 6812.6366 & $6.242 \pm 0.009$ \\
6823.6191  & $6.676 \pm	0.066$ & $6.564	\pm 0.018$ & $2.022 \pm	0.016$ & $0.909	\pm 0.021$ & $0.753 \pm	0.030$ & 6828.64 & $14.664 \pm	0.003$ & $14.013 \pm 0.004$ & 6813.6908 & $6.087 \pm 0.009$ \\
6824.6152  & $6.070 \pm	0.097$ & $6.978	\pm 0.073$ & $1.861 \pm	0.036$ & $0.657	\pm 0.046$ & $0.414 \pm	0.067$ & 6830.64 & $14.708 \pm	0.004$ & $14.063 \pm 0.004$ & 6825.7306 & $7.194 \pm 0.011$ \\
6829.6377  & $7.080 \pm	0.044$ & $9.236	\pm 0.010$ & $2.134 \pm	0.007$ & $0.898	\pm 0.007$ & $0.815 \pm	0.010$ & 6834.64 & $14.642 \pm	0.010$ & $14.005 \pm 0.008$ & 6826.6271 & $7.268 \pm 0.009$ \\
6830.6350  & $7.058 \pm	0.017$ & $8.046	\pm 0.002$ & $2.073 \pm	0.001$ & $0.777	\pm 0.001$ & $0.646 \pm	0.001$ & 6835.64 & $14.650 \pm	0.004$ & $14.037 \pm 0.004$ & 6827.6469 & $6.759 \pm 0.002$ \\
6830.6425  & $7.051 \pm	0.019$ & $8.640	\pm 0.002$ & $2.102 \pm	0.001$ & $0.789	\pm 0.001$ & $0.620 \pm	0.002$ & 6836.64 & $14.656 \pm	0.003$ & $14.042 \pm 0.004$ & 6829.7274 & $7.148 \pm 0.012$ \\
6831.6370  & $6.873 \pm	0.018$ & $7.770	\pm 0.001$ & $1.967 \pm	0.001$ & $0.783	\pm 0.001$ & $0.519 \pm	0.002$ & 6837.64 & $14.706 \pm	0.004$ & $14.033 \pm 0.004$ & 6834.6433 & $7.200 \pm 0.010$ \\
6834.6380  & $7.200 \pm	0.040$ & $9.407	\pm 0.008$ & $1.873 \pm	0.005$ & $0.670	\pm 0.005$ & $0.781 \pm	0.008$ & 6838.64 & $14.647 \pm	0.004$ & $14.032 \pm 0.004$ &           & $            $ \\
6835.6333  & $7.132 \pm	0.016$ & $8.467	\pm 0.002$ & $1.963 \pm	0.001$ & $0.838	\pm 0.001$ & $0.982 \pm	0.001$ & 	  & $	             $ & $		  $ &           & $            $ \\
6836.6370  & $7.243 \pm	0.015$ & $8.216	\pm 0.001$ & $2.088 \pm	0.001$ & $0.896	\pm 0.001$ & $1.034 \pm	0.001$ & 	  & $		     $ & $		  $ &           & $            $ \\
6837.6348  & $7.082 \pm	0.040$ & $8.904	\pm 0.005$ & $2.139 \pm	0.006$ & $0.874	\pm 0.006$ & $1.044 \pm	0.009$ & 	  & $		     $ & $		  $ &           & $            $ \\
6838.6364  & $7.497 \pm	0.013$ & $8.825	\pm 0.002$ & $2.117 \pm	0.001$ & $0.913	\pm 0.001$ & $1.284 \pm	0.001$ & 	  & $		     $ & $		  $ &           & $            $

\enddata 
\label{tab:lc}
\tablenotetext{a}{The continuum emission is listed as the flux density
  at $\lambda 5100$\,\AA~$\times (1+z)$ in units of
  $10^{-15}$\,ergs\,s$^{-1}$\,cm$^{-2}$\,\AA$^{-1}$.}
\tablenotetext{b}{Emission-line light curves are listed as the
  integrated flux in units of $10^{-13}$\,ergs\,s$^{-1}$\,cm$^{-2}$}.
\tablenotetext{c}{$V-$band measurements listed here are scaled to
  match the continuum flux for contemporaneous observations.}
\end{deluxetable}
\clearpage
\end{landscape}

\begin{deluxetable}{lccccccc}
\tablecolumns{8}
\tablewidth{0pt}
\tablecaption{Light-Curve Statistics}
\tablehead{
\colhead{Time Series} &
\colhead{$N$} &
\colhead{$\langle T \rangle$} &
\colhead{$T_{\rm median}$} &
\colhead{$\langle F \rangle$\tablenotemark{a}} &
\colhead{$\langle \sigma_F/F \rangle$} &
\colhead{$F_{\rm var}$} &
\colhead{$R_{\rm max}$}\\
\colhead{} &
\colhead{} &
\colhead{(days)} &
\colhead{(days)} &
\colhead{} &
\colhead{} &
\colhead{} &
\colhead{}
}
\startdata

Continuum    & 46  & $1.4 \pm 1.1$ & 1.0  & $6.80 \pm 0.40$ & $0.0025$ & 0.059  & $1.326 \pm 0.022$ \\
$r$          & 26  & $1.8 \pm 1.5$ & 1.0  & $8.36 \pm 0.54$ & $0.0062$ & 0.064  & $1.254 \pm 0.011$ \\
H$\alpha$    & 30  & $1.6 \pm 1.2$ & 1.0  & $7.27 \pm 1.19$ & $0.0019$ & 0.164  & $1.909 \pm 0.009$ \\
H$\beta$     & 30  & $1.6 \pm 1.2$ & 1.0  & $1.74 \pm 0.29$ & $0.0097$ & 0.171  & $1.841 \pm 0.088$ \\
H$\gamma$    & 30  & $1.6 \pm 1.2$ & 1.0  & $0.68 \pm 0.18$ & $0.0320$ & 0.260  & $2.988 \pm 0.695$ \\
\ion{He}{2}  & 30  & $1.6 \pm 1.2$ & 1.0  & $0.67 \pm 0.32$ & $0.0648$ & 0.474  & $9.640 \pm 4.309$

\enddata
\label{tab:lcstats}
\tablenotetext{a}{Continuum flux density is in units of
                  $10^{-15}$\,ergs\,s$^{-1}$\,cm$^{-2}$\,\AA$^{-1}$
                  and emission line fluxes are in units of
                 $10^{-13}$\,ergs\,s$^{-1}$\,cm$^{-2}$.}
\end{deluxetable}

\begin{deluxetable}{lccccccccccc}
\renewcommand{\arraystretch}{1.8}
\tablecolumns{12}
\tablewidth{0pt}
\tabletypesize{\scriptsize}
\tablecaption{Time Lags, Line Widths, and Virial Products}
\tablehead{
\colhead{} &
\multicolumn{3}{c}{Time Lags}  &
\colhead{} &
\multicolumn{2}{c}{Mean Line Width} &
\colhead{} &
\multicolumn{2}{c}{RMS Line Width} &
\colhead{} &
\multirow{2}{*}{$\frac{c \tau_{\rm cent} \sigma_{\rm line,rms}^2}{G}$} \\
\colhead{Line} &
\colhead{$\tau_{\rm cent}$} &
\colhead{$\tau_{\rm peak}$} &
\colhead{$\tau_{\rm JAV}$} &
\colhead{} &
\colhead{$\sigma_{\rm line}$} &
\colhead{FWHM} &
\colhead{} &
\colhead{$\sigma_{\rm line}$} &
\colhead{FWHM } &
\colhead{} &
\colhead{} \\
\colhead{} &
\colhead{(days)} &
\colhead{(days)} &
\colhead{(days)} &
\colhead{} &
\colhead{(km s$^{-1}$)} &
\colhead{(km s$^{-1}$)} &
\colhead{} &
\colhead{(km s$^{-1}$)} &
\colhead{(km s$^{-1}$)} &
\colhead{} &
\colhead{($10^6$\,M$_{\odot}$)} 
}
\startdata

H$\alpha$   & 2.06$^{+1.42}_{-1.31}$  & 2.75$^{+1.00}_{-2.00}$  & $2.44^{+0.15}_{-0.94}$  && $1781 \pm 36$ & $3032 \pm 54$  && $1783 \pm 66$ & $3579 \pm 145$ && $1.28^{+0.89}_{-0.82}$ \\
H$\beta$    & 2.22$^{+1.19}_{-1.61}$  & 2.50$^{+1.25}_{-2.00}$  & $1.45^{+1.09}_{-0.15}$  && $1821 \pm 53$ & $5688 \pm 163$ && $1544 \pm 98$ & $4615 \pm 330$ && $1.03^{+0.57}_{-0.76}$ \\
H$\gamma$   & 2.14$^{+1.09}_{-1.08}$  & 2.00$^{+1.75}_{-1.00}$  & $1.32^{+0.17}_{-0.67}$  && $1378 \pm 55$ & $4015 \pm 101$ && $1600 \pm 54$ & $4231 \pm 604$ && $1.06^{+0.55}_{-0.54}$ \\
\ion{He}{2} & 0.35$^{+2.17}_{-1.66}$  & 0.00$^{+3.00}_{-1.75}$  & $-0.24^{+0.26}_{-0.28}$ && \nodata       & \nodata        && $2201 \pm 102$ & $4204 \pm 882$&& $0.41^{+2.50}_{-1.92}$ \\
$r$         & 0.35$^{+0.47}_{-0.59}$  & 0.00$^{+0.50}_{-0.25}$  & $0.12^{+0.13}_{-0.13}$  && \nodata       & \nodata        && \nodata       & \nodata        && \nodata  \\

\enddata 
\label{tab:lagwidth}
\tablecomments{Time lags are presented in the observed frame, but are
  corrected for cosmological dilation when calculating the virial
  product --- a tiny effect that is much smaller than the
  uncertainties in the case of NGC\,5273 with $z=0.00362$.}

\end{deluxetable}

\end{document}